\documentclass[RNAAS]{aastex631}

\begin{document}

\title{TeV Emission in NGC 4278: Shock Energetics Insufficient, Compact Origin Favored}

\author{Alberto Dom\'inguez}
\affiliation{IPARCOS and Department of EMFTEL, Universidad Complutense de Madrid, E-28040 Madrid, Spain}

\correspondingauthor{A. Dom\'inguez}
\email{alberto.d@ucm.es}

\begin{abstract}
The Large High Altitude Air Shower Observatory recently detected TeV emission from the low-luminosity active galactic nucleus (LLAGN) NGC\,4278. Integral-field spectroscopy with MEGARA revealed ionized outflows on hundred-parsec scales with kinetic power $\dot E_{\rm shock} \sim 10^{38}$ erg s$^{-1}$, well below the TeV luminosity $L_{\gamma} \gtrsim 10^{41}$ erg s$^{-1}$. The implied efficiencies, $\eta \gtrsim 1000$, are far above unity, showing that shocks cannot account for the TeV emission. Contemporaneous \textit{Fermi}-Large Area Telescope and Swift data, consistent with a compact synchrotron self-Compton jet component, favor a nuclear origin. NGC\,4278 is the first low-ionization nuclear emission-line region (LINER)/LLAGN of its class detected at TeV energies, a benchmark for LINERs with outflows and a prime target for the Cherenkov Telescope Array Observatory. I conclude that the TeV photons arise from a compact nuclear accelerator, while optical outflows trace larger-scale jet--ISM interactions that coexist but do not power the TeV radiation.
\end{abstract}

\keywords{Active galactic nuclei (16) --- Gamma-ray sources (633) --- Radio jets (1347)}

\section{1. Introduction}
Low-luminosity Active Galactic Nuclei (LLAGN) are characterized by inefficient accretion and weak jets \citep{Ho2008}.
The Large High Altitude Air Shower Observatory (LHAASO; \citealt{LHAASO2023}) recently detected TeV emission from NGC\,4278, a nearby elliptical galaxy hosting a low-ionization nuclear emission-line region (LINER)-like nucleus \citep{Giroletti2005, Yuan2009}, revealing very-high-energy (VHE, $\geq 100$ GeV) photons from a system with low-power jets \citep{Cao2024}. This makes NGC\,4278 the first LINER/LLAGN of its class detected at TeV energies. MEGARA integral-field spectroscopy at the 10.4-m Gran Telescopio Canarias (GTC) revealed ionized outflows on scales of $\sim100$~pc aligned with the radio jet \citep{HermosaMunoz2023}. Here I present a new calculation that directly compares the shock energetics with the TeV luminosity to test whether these shocks could provide the required power. In general, three scenarios have been considered for VHE production in LLAGN: (i) extended jet--ISM shocks \citep[e.g.,][]{Bykov2018}, (ii) compact jet zones producing inverse-Compton emission \citep[e.g.,][]{Abdalla2018}, and (iii) magnetospheric gap accelerators near the black hole \citep[e.g.,][]{Aharonian2006}. 
The case of NGC\,4278 allows me to evaluate the relative plausibility of these possibilities.

\section{2. Method}

I computed the $\gamma$-ray luminosity by integrating the LHAASO spectrum:
\begin{equation}
L_{\gamma} = 4\pi d^2 \int_{E_{1}}^{E_{2}} E \, \phi(E)\, dE,
\end{equation}
where $d$ is the luminosity distance, taken as 16.4~Mpc \citep{Tonry2001}, $\phi(E)$ is the differential photon flux, and $E_{1}$ and $E_{2}$ define the energy integration limits. For this purpose, I used the dedicated LHAASO analysis of NGC\,4278 \citep{Cao2024}, which provides the photon flux and photon index. I evaluated the luminosity in the 1–10~TeV band, directly constrained by the data. Attenuation by the extragalactic background light (EBL) up to 10~TeV at the distance of NGC\,4278 implies that the intrinsic luminosities in these bands are slightly higher than observed, with corrections of at most 10\% \citep[e.g.,][]{Saldana-Lopez2021,Dominguez2024}.

The shock kinetic power was estimated from the MEGARA observations using
\begin{equation}
\dot{E}_{\mathrm{shock}} = \frac{1}{2} \, \dot{M} \, \left( v^{2} + 3\sigma^{2} \right),
\end{equation}
where $\dot M$ is the ionized gas mass outflow rate, estimated to be $(4\pm 1)\times10^{-3}$~M$_\odot$~yr$^{-1}$, $v$ is the outflow velocity, of order a few 100~km~s$^{-1}$, and $\sigma$ is the velocity dispersion of the outflowing ionized gas, which is $(49\pm 27)$~km~s$^{-1}$ \citep{Rose2018,HermosaMunoz2023}. The outflow velocity measures the bulk motion of the gas as it flows outward, while the velocity dispersion quantifies the random or turbulent spread of velocities that broadens the emission line around that bulk motion. These values are derived from spatially resolved spectroscopy and include the uncertainties from line diagnostics, geometry, and extinction corrections \citep{HermosaMunoz2023}.

\section{3. Results}
I find an observed TeV luminosity of $L_{1-10\,\mathrm{TeV}}\sim 10^{41}$ erg s$^{-1}$, based on the measured photon flux $F_{1-10\,\mathrm{TeV}}^{\rm ph}= (7.0\pm 1.1_{\mathrm{sta}} \pm 0.35_{\mathrm{sys}})\times10^{-13}$ cm$^{-2}$ s$^{-1}$ and photon index $\Gamma=2.56\pm 0.14$, obtained by integrating the power-law fit over 1--10 TeV \citep{Cao2024}. This value should be regarded as a lower limit, since the TeV emission may extend beyond the measured range, contributing additional power. In comparison, the shock power is only $\dot E_{\rm shock} \sim 10^{38}$~erg~s$^{-1}$, around three order of magnitude below the TeV luminosity (see Table~\ref{tab1}). The putative efficiency required to power the TeV emission is therefore
\begin{equation}
\eta = \frac{L_{\gamma}}{\dot E_{\rm shock}} \gtrsim 1000,
\end{equation}
For comparison, efficiencies inferred for shock-powered nonthermal emission in other LLAGN and in supernova-driven galactic winds are typically well below unity \citep[e.g.,][]{Abdalla2018, Bykov2018}. The efficiency values required in NGC\,4278 are unphysical, thereby excluding extended jet--ISM interactions as the TeV source. Note that accounting for EBL attenuation further strengthens this result. Such a direct energy-budget comparison has not been previously presented for NGC\,4278.

\begin{table}[h!]
\caption{Comparison of the observed TeV luminosity and shock energetics in NGC\,4278.}
\label{tab1}
\centering
\begin{tabular}{lcc}
\hline\hline
Quantity & Value & Reference \\
\hline
$L_{\gamma}$ (1--10 TeV) & $0.8\times10^{41}$ erg s$^{-1}$ & \citet{Cao2024} \\
$\dot E_{\rm shock}$ & $(7.2\pm 1.7)\times 10^{37}$ erg s$^{-1}$ & \citet{HermosaMunoz2023} \\
\hline
\end{tabular}
\end{table}

\section{4. Discussion and Conclusions}

My calculation suggests that extended shocks are not sufficient to account for the TeV emission from NGC\,4278. 
Additional evidence comes from \citet{Bronzini2024}, who reported a contemporaneous \textit{Fermi}-LAT detection with a very hard spectrum and a nuclear high state in Swift observations. This behavior is consistent with a compact synchrotron self-Compton jet zone and disfavors a contribution from magnetospheric gap acceleration \citep[e.g.,][]{Aharonian2006}. The case is further strengthened by the observed month-scale variability, which implies an emitting region of size $R \lesssim 10^{-3}$~pc, far smaller than the $\sim100$~pc-scale outflows. Other interpretations, including shock-related scenarios and hadronic models, have also been recently proposed \citep{Lian2024,Shoji2025}. Taken together, these results support a scenario in which the compact nuclear region produces the VHE emission, while the extended shocks seen with MEGARA trace the interaction of the jet with the surrounding medium but do not power the TeV radiation.

I note that the LHAASO detection and the MEGARA optical spectroscopy were not contemporaneous. The LHAASO measurement, however, corresponds to an average over an extended observing campaign, while the contemporaneous \textit{Fermi}-LAT and \textit{Swift} data reported by \citet{Bronzini2024} reveal enhanced nuclear activity with a hard GeV spectrum and an X-ray high state, supporting a compact origin. My comparison therefore assumes that the large-scale outflows and the nuclear TeV activity are persistent features rather than isolated transients. This assumption is justified since the ionized outflows extend over $\sim100$~pc and have dynamical timescales of $10^5$--$10^6$ years, making them effectively steady on observational timescales. The nuclear TeV activity, while variable on month-scale intervals, is consistent with a compact accelerator capable of sustaining recurrent emission episodes over long periods. Thus, the lack of contemporaneity between the optical and TeV data does not compromise the validity of the energy-budget comparison presented here.

More broadly, NGC\,4278 may be representative of a wider class of LINERs with ionized outflows. Recent surveys indicate that six out of nine well-studied LINERs host outflowing gas on hundred-parsec scales \citep{HermosaMunoz2023}, yet none except NGC\,4278 has been firmly detected at TeV energies so far. This makes NGC\,4278 a benchmark case to evaluate whether such systems can also act as VHE emitters. Its detection shows that TeV production is not restricted to blazars or powerful Fanaroff–Riley type I radio galaxies, but extends to the LINER/LLAGN population, with NGC\,4278 providing a benchmark for LINERs with outflows and prospective Cherenkov Telescope Array Observatory targets.

\begin{acknowledgments}
I thank Vaidehi Paliya and Armando Gil de Paz for helpful discussions.

\end{acknowledgments}

\end{document}